\documentclass[amsmath,prb,twocolumn,showpacs,superscriptaddress]{revtex4}

\usepackage{amssymb}
\usepackage{graphicx}
\usepackage{color}

\begin{document}

\title{Nuclear dynamics at molecule-metal interfaces: A pseudoparticle perspective}

\author{Michael Galperin}
\affiliation{Department of Chemistry and Biochemistry, University of California at San Diego, La Jolla, CA 92093, USA}
\author{Abraham Nitzan}
\affiliation{Department of Chemistry, University of Pennsylvania, Philadelphia, Pennsylvania 19104, USA}
\affiliation{School of Chemistry, Tel Aviv University, Tel Aviv, 69978, Israel}

\begin{abstract}
We discuss nuclear dynamics at molecule-metal interfaces including non-equilibrium molecular junctions. Starting from the many-body states (pseudoparticle) formulation of the molecule-metal system in the molecular vibronic basis, we introduce gradient expansion in order to reduce the adiabatic nuclear dynamics (that is, nuclear dynamics on a single molecular potential surface) into its semi-classical form while maintaining the effect of the non-adiabatic electronic transitions between different molecular charge states. This yields a set of equations for the nuclear dynamics in the presence of these non-adiabatic transitions, which reproduce surface hopping formulation in the limit of small metal-molecule coupling (where broadening of the molecular energy levels can be disregarded) and Ehrenfest dynamics (motion on the potential of mean force) 
when information on the different charging states is traced out, which is relevant 
when this coupling is strong.
\end{abstract}


\maketitle


\section{Introduction}
The coupled electronic-nuclear dynamics in molecules positioned at and interacting 
with metal interfaces presents a fundamental problem that stems from the fact that 
in such systems the usual timescale separation between electron and nuclear dynamics 
does not necessarily holds.  The problem has gained renewed interest 
in the context of nuclear dynamics in molecular conduction junctions, a presently
active field of research due to its fundamental and applicational importance.\cite{LorenteJPCM05,MGRatnerNitzanJPCM07,GalperinRatnerNitzanTroisiScience08,SeidemanAccChemRes10,BaowenHanggiRMP12}
Experimental measurements of inelastic electron tunneling spectroscopy,\cite{McEuenNature00,HoPRL00,ZhitenevMengBaoPRL02,ReedNL04,DekkerNature04}
and more recently fluorescence\cite{ZhangNL07,HoPRB08,HoPRL10,SchullPRL14} 
and Raman\cite{CheshnovskySelzerNatNano08,NatelsonNL08,NatelsonNatNano11,ApkarianACSNano12,ApkarianHessNL13}  
spectroscopies serve as tools capable of providing information on presence of 
the molecule in the junction and extent of heating of the device. 
Description of transport,\cite{ReppNatPhys10} 
heating,\cite{KochVonOppenNitzanPRB06} instabilities,\cite{LiljerothReppMeyerScience07,SegalPCCP12} and current (and light) 
induced chemistry\cite{SeidemanJPCM03,SeidemanAccChemRes10,ReppPerssonPRL10} 
in junctions often require quantum-mechanical description beyond the Born-Oppenheimer
approximation. 

In junctions, electron transition events between molecule and contacts
result in coupling between different adiabatic potential
surfaces, resulting in non-adiabatic molecular dynamics (NAMD). 
NAMD plays important role in many chemical dynamics processes, ranging from 
surface chemistry to spectroscopy, radiationless electronic relaxation,
photochemistry, and electron transfer.\cite{Tully_review_JPCC09}
Considerations  of non-adiabaticity are particularly important fro molecules
that exchange electrons with metal or semiconductor substrates
because the rate of this exchange can be smaller or larger than characteristic nuclear
timescales. Consequently, NAMD can drastically influences the response of molecular junctions, 
and can dominate the transport behavior associated with many interesting phenomena 
ranging from current induced chemistry to molecular motors.\cite{SeidemanJPCM03,TodorovNatNano09}

Full quantum-mechanical solution of electron-nuclear dynamics is possible only 
for relatively small systems.\cite{WhiteGalperinPCCP12,RabaniPRB13}  
Thus one has to rely on quasi-classical formulations.\cite{StockThossAdvChemPhys05,MillerJCP98}
Among them Ehrenfest dynamics\cite{NewnsPRB95,TodorovJPCM04,TodorovJPCM05,TodorovPRB12} and fewest switches surface hopping (SH) algorithm\cite{TullyJCP90,TullyJCP95} 
are employed most often. 
The latter was applied to many problems in the gas phase,\cite{TretiakMozyrskyNatCommun13,TretiakJCP13,WhiteTretiakMozyrskyJCP14,WhiteTretiakMozyrskyJCP15}  and recently also 
to molecules near metallic surfaces.\cite{Tully_review_JPCC09,TullyJCP09,TullyJCP09a,NitzanSubotnikJCP15,SubotnikJCP15,NitzanSubotnikJCP15_2,NitzanSubotnikJCP15_3} 
From theoretical perspective, the Ehrenfest method can be obtained as an expansion around
the stationary (classical) solution of the quantum electron-nuclei problem.\cite{NewnsPRB95,TodorovPRB12}
Originally surface hopping algorithm  was formulated in an ad hoc manner.\cite{TullyJCP90}
Later work has discussed its relation to the quantum-classical Liouville equation.\cite{KapralAnnRevPhyChem06,KapralJCP07,SubotnikJCP13}
Such considerations are not readily suitable for molecule-metal systems that
are characterized by frequent exchange of electrons between a molecule
and an electronic continuum as well as broadening of the molecular levels. 
Here we focus on this type of systems.

In the absence of molecule-metal interaction, the molecule is presented in terms of 
its many-body molecular states that are usually described within 
the Born Oppenheimer (BO) approximation. The latter is based on the assumption 
that nuclei are slow relative to the electronic dynamics.   
In the other extreme limit of strong molecule-metal interaction, 
where the molecule-metal electron exchange is also fast relative to the nuclear dynamics, 
the BO approximation is set with respect to hybrid molecule-metal  electronic 
states.\cite{GalperinNitzanRatnerARXIV09}  
In the intermediate situation of weak but non-vanishing molecule-metal electron 
exchange coupling, the BO approximation breaks down and the system dynamics 
includes transitions between electronic states of different charges that take place 
on timescale of the nuclear dynamics. The ensuing dynamics can be described in 
the basis of the BO states of the isolated molecule or in the BO states of the strongly 
coupled molecule-metal complex by incorporating surface hopping (SH) events 
into the corresponding nuclear dynamics. In either case, the equations of motion 
used to describe the mixed quantum-classical dynamics have been postulated 
rather than derived.  In the present communication we offer a systematic derivation 
of the equations of motion for such systems, and discuss their limiting behaviors: 
a surface hopping algorithm in the limit of weak molecule-metal coupling and 
Ehrenfest dynamics on the potential of mean force in the limit where the electron 
exchange rate exceeds the characteristic nuclear dynamics.   

Our starting point is the observation that the molecular process under discussion is
electron transfer into and out of the molecule which is most naturally described in
the language of many-body molecular (here vibronic) states. 
This in turn requires an appropriate formulation of transport in the same language.
The goal of this paper is to present such a derivation,
which starts from the full quantum-mechanical description, and step-by-step
derives equations suitable for implementation of the surface hopping algorithm 
for non adiabatic molecular dynamics at molecule-metal interfaces. 
The presented derivation extends recent considerations\cite{SubotnikJCP15,NitzanSubotnikJCP15_2,NitzanSubotnikJCP15_3,NitzanSubotnikJCP15} 
by taking into account hybridization (broadening) of molecular states with those 
of the metal(s) in a rigorous way
and by providing expressions suitable for implementation of the algorithm in 
current carrying molecular junctions.

The structure of the paper is as follows: after introducing the model in Section~\ref{model} 
we shortly discuss (Section~\ref{method}) the pseudoparticle non-equilibrium Green's 
function (PP-NEGF) methodology, which allows to formulate the molecular junction problem 
in the language of many-body states, and apply it to formulate 
exact equations-of-motion (EOMs) for the electron-nuclei model. 
Next, in Section~\ref{grad} we consider the first order gradient expansion of these equations,
which casts the nuclear dynamics in a classical form while maintaining the effect of 
the non-adiabatic electronic transitions on this dynamics.
This yields a general formulation of the non-adiabatic dynamics at molecule-metal 
interfaces with both optical (intra-molecular) and charge transfer events present. 
The resulting semiclassical EOMs can be used as a basis for the surface hopping 
treatment of non-adiabatic dynamics in junctions. 
We specialize to the simple model of a resonant level coupled linearly  
to a single vibration in Section~\ref{osc} in order to discuss connection to previous work. 
Section~\ref{conclude} concludes.

\section{Model}\label{model}
We start form the usual representation of the system
where both electron ($e$) and nuclear ($n$) dynamics is taken into account. 
The Hamiltonian is
\begin{equation}
\label{H_FQ}
\hat H(r,R) = \hat T_n(R) + \hat V_{nn}(R) + \hat T_e(r)+\hat V_{ee}(r)+\hat V_{en}(r,R)
\end{equation}
where $r$ and $R$ stand for the coordinates of all electrons and all nuclei in the system,
respectively. $\hat T_n$ ($\hat T_e$) is the kinetic energy of the nuclei (electrons) 
and $\hat V_{nn}$ ($\hat V_{ee}$) is the Coulomb interaction between nuclei (electrons),
while $\hat V_{en}$ is the electron-nuclear attraction. Explicit expressions are
\begin{align}
\label{Tn}
\hat T_n(R) =& -\sum_{a=1}^{N_n}\frac{1}{2M_a}\Delta_{\vec R_a}
\\
\label{Vnn}
\hat V_{nn}(R) =& \sum_{a,b=1}^{N_n}\frac{Z_aZ_b}{\lvert\vec R_a-\vec R_b\rvert}
\\
\label{Te}
\hat T_e(r) =& -\sum_{i=1}^{N_e}\frac{1}{2}\Delta_{\vec r_i}
\\
\hat V_{ee}(r) =& \sum_{i,j=1}^{N_e}\frac{1}{\lvert\vec r_i-\vec r_j\rvert}
\\
\label{Ven}
\hat V_{en}(r,R) =& -\sum_{a=1}^{N_a}\sum_{i=1}^{N_e}\frac{Z_a}{\lvert\vec r_i-\vec R_a\rvert}
\end{align}
Here $N_a$ and $N_e$ represent the total numbers of atoms and electrons in the system,
respectively. Here and below we have utilized atomic units, i.e. $m_e=k_b=\hbar=1$.

Our goal is to describe electronic and nuclear dynamics in a 
model junction that consists of a molecule $M$ coupled to a number 
of metallic contacts $K$. The latter are free electron reservoirs each at its own equilibrium 
(i.e. characterized by temperature $T_K$ and electrochemical potential $\mu_K$). 
To do so we (artificially) separate the whole system into molecular and contacts parts
and assume that their electronic structure has been determined.
Nuclear dynamics is assumed to be confined to the molecular region only 
(and from now on we reserve $R$ to represent the coordinates of the molecular atoms) 
with the contacts atoms treated as static. Coupling between molecule and contacts is taken
(as usual) to be single-particle operator (i.e. electron-electron interaction between
electrons in $M$ and $K$ is disregarded).
Below we take the index $k$ to indicate both the band and the wave vector of an electron 
and use the second quantized representation of these states. 
The molecular subsystem will be treated in the language of vibronic states,
which can be expanded in the basis of Born-Oppenheimer states\cite{Nitzan_2006}\footnote{Note in passing that alternatively exact states of molecular system\cite{GrossPhilTransRSocA14} 
can be used as a basis.}
\begin{equation}
\label{BO}
\Phi_{ev}(r,R)=\psi_e(r,R)\,\chi_v^e(R)\equiv\lvert e\, v\rangle
\end{equation}
This yields the junction Hamiltonian in a mixed representation,
where the molecule is described in terms of its vibronic states while 
the contacts are represented in the single-electron second quantized form,
\begin{equation}
\label{H}
\hat H = \hat H_M + \sum_K\left(\hat H_K +\hat V_K\right)
\end{equation}
where 
\begin{align}
\hat H_M=& \sum_{e_1v_1,e_2v_2\in M} 
   H^M_{e_1v_1,e_2v_2}\hat X_{e_1v_1,e_2v_2}
\\
\hat H_K=& \sum_{k\in K} \varepsilon_k\hat c_k^\dagger\hat c_k
\\
\hat V_K=& \sum_{k\in K}\sum_{e_1v_1,e_2v_2\in M} 
 V^K_{k,(e_1v_1,e_2v_2)} \hat c_k^\dagger \hat X_{e_1v_1,e_2v_2} + H.c.
\end{align}
where $c_k^\dagger$ ($\hat c_k$) creates (annihilates) an electron in level $k$ of the contacts,
$\hat X_{e_1v_1,e_2v_2}\equiv \lvert e_1v_1\rangle\langle e_2v_2\rvert$ is 
the molecular Hubbard (projection) operator, and
\begin{align}
\label{HM_ME}
&  H^M_{e_1v_1,e_2v_2} = \langle e_1v_1\rvert
\hat H_M
\lvert e_2v_2\rangle
\\
& \equiv \int dr\int dR\, \overset{*}{\Phi}{}_{e_1v_1}(r,R)\,\hat H_M(r,R)\,\Phi_{e_2v_2}(r,R)
\nonumber \\
\label{VK_ME}
& V^K_{k,(e_1v_1,e_2v_2)} = \sum_{i=1}^{N_{e_2}}\int dr_M\int dR\,
\overset{*}{\psi}{}_k(\vec r_i)\,\overset{*}{\Phi}{}_{e_1v_1}(r/ \vec r_i,R)
\nonumber \\ &\qquad\qquad\qquad \times
\hat O_1(r,R)\,\Phi_{e_2v_2}(r,R)
\end{align} 
are matrix elements for the molecular Hamiltonian and coupling to contact $K$.
Here $\int dr_M\ldots$ integrates over electrons on the molecule,
$N_{e_2}$ is number of electrons in the state $\lvert e_2v_2\rangle$,  
$\Phi_{e_1v_1}(r/ \vec r_i,R)$ indicates vibronic state $\lvert e_1v_1\rangle$ 
with one electron, $\vec r_i$, less than in the state $\lvert e_2v_2\rangle$, 
and $\hat O_1(r,R)$ is a single-electron operator, 
which (depending on the problem) can include contributions from 
(\ref{Te}) or (\ref{Ven}).

\section{Method}\label{method}
Evaluating the dynamics of systems described by Hamiltonians of the type of Eq.(\ref{H})
in terms of the many-body states of the isolated system,
{\em the nonequilibrium atomic limit},\cite{WhiteOchoaMGJPCC14}
can be treated within a number of techniques. 
Among them are the generalized quantum master equation,\cite{PedersenWackerPRB05,EspGalpPRB09,EspositoMGJPCC10,LeijnseWegewijsPRB08,GrifoniLeijnseWegewijsPRB10,WegewijsPRB12,WegewijsPRB14} 
projection operator,\cite{OchoaMGRatnerJPCM14}
Hubbard\cite{SandalovIJQC03,SandalovJPCM06,SandalovPRB07,FranssonPRB05,MGNitzanRatnerPRB08,YeganehNL09} 
and pseudo particle (PP)\cite{WingreenMeirPRB94,EcksteinWernerPRB10,OhAhnBubanjaPRB11,WhiteGalperinPCCP12,WhiteFainbergMGJPCL12,WhiteMiglioreMGANJCP13,WhiteTretiakNL14,WernerRMP14}
nonequilibrium Green's functions (NEGF) formulations, 
numerically exact renormalization group approaches\cite{Schoeller2000,WhitePRB05,DagottoPRB08,UedaJPhysSocJpn08,UedaAnnDerPhys11}
and quantum Monte Carlo methodologies.\cite{RabaniPRL08,SegalMillisReichmanPRB10,CohenRabaniPRB11,CohenRabaniPRB13,RabaniPRB13,SegalJCP13}
The latter is usually too heavy to be utilized in realistic simulations.

Here we use the PP-NEGF methodology in the lowest order (non-crossing) 
approximation (NCA).
We note that generalization to higher orders is straightforward.\cite{EcksteinWernerPRB10}
The PP-NEGF formulation is based on the introduction of second quantization in the space of
the many-body system states
\begin{equation}
\label{pev}
\lvert ev\rangle = \hat p_{ev}^\dagger\lvert 0\rangle
\end{equation}
where $\lvert 0\rangle$ is vacuum state. 
The creation, $\hat p_{ev}^\dagger$, and annihilation, $\hat p_{ev}$, operators 
satisfy the usual commutation relations of either Fermi or Bose operators 
depending on the number of electrons in the state $\lvert ev\rangle$
This formulation generates an extended Hilbert space,
in which the physical subspace is defined by the normalization condition
\begin{equation}
\label{norm}
 \sum_{ev} \hat p_{ev}^\dagger\hat p_{ev}=1
\end{equation}
The dynamical evolution of the system is expressed in terms of the
{\em pseudoparticle Green function}, defined on the Keldysh contour as
\begin{equation}
\label{Gev}
G_{e_1v_1,e_2v_2}(\tau_1,\tau_2) \equiv
   -i\langle T_c\, \hat p_{e_1v_1}(\tau_1)\,\hat p_{e_2v_2}^\dagger(\tau_2)\rangle
\end{equation}
where $T_c$ is the contour ordering operator and $\tau_{1,2}$ are the contour variables.
For our consideration it is convenient to represent this Green function (GF) in
a different basis as follows
\begin{align}
\label{GeR}
&G_{e_1,e_2}(R_1,\tau_1;R_2,\tau_2)\equiv
\\ &\qquad\sum_{v_1,v_2}
\chi_{v_1}^{e_1}(R_1)\, G_{e_1v_1,e_2v_2}(\tau_1,\tau_2) \,
\overset{*}{\chi}{}_{v_2}^{e_2}(R_2)
\nonumber
\end{align}
where $\chi_v^e(R)$ is the vibrational wavefunction of the BO approximation (\ref{BO})
in the isolated molecule.
The retarded projection of the GF (\ref{GeR}), $G^r_{e_1,e_2}(R_1,t_1;R_2,t_2)$, 
gives information on the many-body spectral function of the system
\begin{align}
\label{defA}
 &A_{e_1e_2}(R_1,t_1;R_2,t_2)=
 \\ &
 i\left( G^r_{e_1,e_2}(R_1,t_1;R_2,t_2)
 - G^a_{e_1,e_2}(R_1,t_1;R_2,t_2) \right)
 \nonumber
\end{align}
where  $G^a_{e_1,e_2}(R_1,t_1;R_2,t_2) \equiv \overset{*}{G}{}^r_{e_2,e_1}(R_2,t_2;R_1,t_1)$,
while its lesser projection, $G^{<}_{e_1,e_2}(R_1,t_1;R_2,t_2)$, contains information
on nonequilibrium distribution in the many-body states space of the molecule.
These projections satisfy the usual Dyson equation. 
In particular, the following expressions are exact 
\begin{widetext}
\begin{align}
 \label{GltEOM}
 & i\left(\frac{\partial}{\partial t_1} + \frac{\partial}{\partial t_2}\right) G^{<}_{e_1,e_2}(R_1,t_1;R_2,t_2)
 +\sum_e\bigg(G^{<}_{e_1,e}(R_1,t_1;R_2,t_2)\hat H^M_{e,e_2}(R_2)
 -\hat H^M_{e_1,e}(R_1)G^{<}_{e,e_2}(R_1,t_1;R_2,t_2)\bigg)
\nonumber \\ &
 =\sum_e\int dR\int ds\bigg(\Sigma^{<}_{e_1,e}(R_1,t_1;R,s)G^a_{e,e_2}(R,s;R_2,t_2)
 +\Sigma^{r}_{e_1,e}(R_1,t_1;R,s)G^{<}_{e,e_2}(R,s;R_2,t_2)
 \\ &\qquad\qquad\qquad\quad
 -G^{<}_{e_1,e}(R_1,t_1;R,s)\Sigma^{a}_{e,e_2}(R,s;R_2,t_2)
 -G^{r}_{e_1,e}(R_1,t_1;R,s)\Sigma^{<}_{e,e_2}(R,s;R_2,t_2)\bigg)
 \nonumber \\
 \label{GrEOM}
& i\left(\frac{\partial}{\partial t_1} - \frac{\partial}{\partial t_2}\right) G^{r}_{e_1,e_2}(R_1,t_1;R_2,t_2) 
-\sum_e\bigg(G^{r}_{e_1,e}(R_1,t_1;R_2,t_2)\hat H^M_{e,e_2}(R_2)
 +\hat H^M_{e_1,e}(R_1)G^{r}_{e,e_2}(R_1,t_1;R_2,t_2)\bigg)
\nonumber \\ &
=\delta_{e_1,e_2}\,\delta(R_1-R_2)\,\delta(t_1-t_2)
\\ &\qquad
+\sum_e\int dR\int ds\bigg(
\Sigma^{r}_{e_1,e}(R_1,t_1;R,s)G^{r}_{e,e_2}(R,s;R_2,t_2)
+G^{r}_{e_1,e}(R_1,t_1;R,s)\Sigma^{r}_{e,e_2}(R,s;R_2,t_2)\bigg)
\nonumber
\end{align}
Here 
\begin{equation}
\label{HMe1e2}
 \hat H^M_{e_1,e_2}(R)\equiv \int dr\, \overset{*}{\psi}{}_{e_1}(r,R)\, \hat H_M(r,R)\, \psi_{e_2}(r,R)
 = \delta_{e_1,e_2}\bigg(\hat T_n(R) + \hat V_{e}(R)\bigg) + \hat d_{e_1,e_2}(R)
 + \hat f_{e_1,e_2}(R) 
\end{equation}
where $\hat T_n(R)$ is defined in (\ref{Tn}), $\hat V_e(R)\equiv\hat V_{nn}(R)+E_e(R)$ is 
the adiabatic surface ($\hat V_{nn}(R)$ is defined in (\ref{Vnn})), 
$E_e(R)$ is the electron eigenenergy: $\big(\hat T_e(r)+\hat V_{ee}(r)+\hat V_{ne}(r,R)\big)\psi_e(r,R)=E_e(R)\psi_e(r,R)$,
and 
\begin{align}
 \hat d_{e_1,e_2}(R)\equiv& -\sum_{a=1}^{N_a}\frac{1}{M_a}
 \int dr\, \overset{*}{\psi}{}_{e_1}(r,R)\frac{\partial\psi_{e_2}(r,R)}{\partial\vec R_a}\,
 \frac{\partial}{\partial\vec R_a}
 \equiv \sum_{a=1}^{N_a} \vec d^a_{e_1,e_2}(R)\, \frac{\partial}{\partial\vec R_a}
 \\
 f_{e_1,e_2}(R) \equiv& -\sum_{a=1}^{N_a}\frac{1}{2M_a}\int dr\,
 \overset{*}{\psi}{}_{e_1}(r,R)\,\frac{\partial^2\psi_{e_2}(r,R)}{\partial\vec R_a^2}
\end{align}
are the intra-molecular  (not related to electron transfer between molecule and contacts) 
non-adiabatic couplings.  Note that these will not couple between states of different charges. 
$\Sigma^{<,r,a}_{e_1,e_2}(R_1,t_1;R_2,t_2)$ in (\ref{GltEOM}) and (\ref{GrEOM}) 
are the lesser, retarded, and advanced projections of self-energy due to coupling to metallic contacts. 
Explicit expression for the latter within the NCA is\cite{WhiteGalperinPCCP12}
\begin{align}
 \Sigma_{e_1,e_2}(R_1,\tau_1;R_2,\tau_2)=&
i\sum_{e_1',e_2'}\int dR_1'\int dR_2' \,
G_{e_1',e_2'}(R_1',\tau_1;R_2',\tau_2)
\\ & \times
\sum_K\bigg(c^K_{e_1,e_1';e_2,e_2'}(R_1,R_1',\tau_1;R_2,R_2',\tau_2)
-c^K_{e_2',e_2;e_1',e_1}(R_2',R_2,\tau_2;R_1',R_1,\tau_1)\bigg)
\nonumber
\end{align}
\end{widetext}
where $c^K_{e_1,e_1';e_2,e_2'}(R_1,R_1',\tau_1;R_2,R_2',\tau_2)$ is the
correlation between two electron transitions from the bath to the system:
one at time $\tau_1$ with molecular electronic state going from
$e_1'\to e_1$ and nuclei changing their positions from $R_1'\to R_1$,
the other at time $\tau_2$ with molecular electronic state undergoing transformation from  
$e_2'\to e_2$ with nuclei moving $R_2'\to R_2$. 
\begin{align}
& c^K_{e_1,e_1';e_2,e_2'}(R_1,R_1',\tau_1;R_2,R_2',\tau_2) \equiv
\nonumber \\ &
\sum_{v_1,v_1',v_2,v_2'}
\overset{*}{\chi}{}_{v_1}^{e_1}(R_1)\,\chi_{v_1'}^{e_1'}(R_1')\,
\overset{*}{\chi}{}_{v_2'}^{e_2'}(R_2')\,\chi_{v_2}^{e_2}(R_2)
\\ &\times
\sum_{k\in K}V^K_{(e_1'v_1',e_1v_1),k}\, g_k(\tau_1,\tau_2)\,V^K_{k,(e_2'v_2',e_2v_2)}
\nonumber
\end{align}
where $g_k(\tau_1,\tau_2)\equiv -i\langle T_c\,\hat c_k(\tau_1)\,\hat c_k^\dagger(\tau_2)\rangle$
is the Green's function of free electron in state $k$ of the contact $K$. 

\section{Gradient expansion}\label{grad}
Assuming slow nuclear dynamics we perform first order gradient expansion  
with respect to time and nuclear coordinates, keeping the electronic dynamics as purely quantum. 
Starting with the PP-NEGF EOMs (\ref{GltEOM}) and (\ref{GrEOM}) allows to keep
information on the potential energy surface while going to quasi-classical description
of the nuclear motion. Following the standard procedure\cite{HaugJauho_2008}
we transfer to the Wigner variables, introducing slow (classical) coordinates and time
\begin{equation}
 R=\frac{R_1+R_2}{2}\qquad t=\frac{t_1+t_2}{2}
\end{equation}
and fast (quantum) variables
\begin{equation}
 R_q=R_1-R_2\qquad t_q=t_1-t_2,
\end{equation}
so that $f(R_1,t_1;R_2,t_2)\to f(R,t;R_q,t_q)$ ($f$ is an arbitrary correlation function),
and perform Fourier transform in the latter 
\begin{equation}
f(R,t;p,E)\equiv \int dR_q\int dt_q\, e^{-ipR_q+iEt_q}\, f(R,t;R_q,t_q)
\end{equation}
\begin{widetext}
Performing first order gradient expansion in Eq.(\ref{GltEOM}) leads to
\begin{align}
\label{Gltgrad1}
\frac{\partial}{\partial t}G^{<}_{e_1,e_2}(R,t;p,E)=&
-i\sum_{e_3,e_4}\hat{\mathcal{L}}{}_{e_1,e_2;e_3,e_4}(R,p)\, G^{<}_{e_3,e_4}(R,t;p,E)
\\ &
-\sum_{e_3,e_4}\int dR' \int dp'\int \frac{dE'}{2\pi}\, 
\sum_K\hat{\mathcal{D}}{}^{K}_{e_1,e_2;e_3,e_4}(R,R',t;p,p',E,E')\, G^{<}_{e_3,e_4}(R',t;p',E')
\nonumber
\end{align} 
where 
\begin{align}
& \hat{\mathcal{L}}{}_{e_1,e_2;e_3,e_4}(R,p) \equiv  
i\, \delta_{e_1,e_3}\delta_{e_2,e_4}\,
  \bigg[\frac{\partial V_{e_1}(R)}{\partial \vec R}\frac{\partial}{\partial\vec p}
   -\vec p\,\frac{\partial}{\partial\vec R}\bigg]
\nonumber \\ &
 +\delta_{e_2,e_4}\bigg(
 f_{e_1,e_3}(R)+\bigg[i \vec p-\frac{1}{2}\frac{\partial}{\partial\vec R}\bigg] \vec d_{e_1,e_3}(R)
 +\frac{1}{2}\frac{\partial}{\partial\vec R}\bigg[i f_{e_1,e_3}(R)-\vec d_{e_1,e_3}(R)\cdot \vec p\bigg]
 \frac{\partial}{\partial\vec p} +\frac{1}{2}\vec d_{e_1,e_3}(R)\frac{\partial}{\partial\vec R}
 \bigg)
 \\ &
-\delta_{e_1,e_3}\bigg(
 f_{e_4,e_2}(R)-\bigg[i \vec p+\frac{1}{2}\frac{\partial}{\partial\vec R}\bigg] \vec d_{e_4,e_2}(R)
 -\frac{1}{2}\frac{\partial}{\partial\vec R}\bigg[i f_{e_4,e_2}(R)+\vec d_{e_4,e_2}(R)\cdot \vec p\bigg]
 \frac{\partial}{\partial\vec p} +\frac{1}{2}\vec d_{e_4,e_2}(R)\frac{\partial}{\partial\vec R}
 \bigg) 
 \nonumber
\end{align}
is the Liouvillian superoperator of  the free molecular evolution, and
\begin{align}
\label{DK}
&\hat{\mathcal{D}}{}^{K}_{e_1,e_2;e_3,e_4}(R,R',t;p,p',E,E')=
\hat{\mathcal{D}}{}^{K\,(0)}_{e_1,e_2;e_3,e_4}(R,R',t;p,p',E,E')
+\hat{\mathcal{D}}{}^{K\,(1)}_{e_1,e_2;e_3,e_4}(R,R',t;p,p',E,E')
\\
\label{DK0}
&\hat{\mathcal{D}}{}^{K\,(0)}_{e_1,e_2;e_3,e_4}(R,R',t;p,p',E,E')=
\delta(R-R')\,\delta(p-p')\,\delta(E-E')\sum_{e_s,e_s'}\int dR_s\int dp_s\int \frac{dE_s}{2\pi}
\nonumber \\ &
\bigg(
\bigg[ 
\delta_{e_1,e_3} \big(c^{K\, >}_{e_4,e_s';e_2,e_s}(R,R_s,t;p,-p_s,E-E_s)
                                 -c^{K\, <}_{e_s,e_2;e_s',e_4}(R,R_s,t;-p,p_s,E_s-E)\big) 
                            G^a_{e_s',e_s}(R_s,t;p_s,E_s)
\nonumber \\ &
+\delta_{e_2,e_4}\,G^r_{e_s,e_s'}(R_s,t;p_s,E_s)
  \big(c^{K\, <}_{e_s',e_3;e_s,e_1}(R,R_s,t;-p,p_s,E_s-E)
        -c^{K\, >}_{e_1,e_s;e_3,e_s'}(R,R_s,t;p,-p_s,E-E_s)\big)
\bigg]
\\ &
+\sum_e\bigg(
\big(c^{K\, >}_{e_4,e;e_3,e_1}(R,R',t;-p,p',E'-E)-c^{K\, <}_{e_1,e_3;e,e_4}(R,R',t;p,-p',E-E')\big)
      G^a_{e,e_2}(R,t;p,E)
\nonumber \\ &\qquad\quad
+G^r_{e_1,e}(R,t;p,E)
\big(c^{K\, <}_{e,e_3;e_2,e_4}(R,R',t;p,-p',E-E')-c^{K\, >}_{e_4,e_2;e_3,e}(R,R',t;-p,p',E'-E)\big)
\bigg)
\nonumber
\end{align}
is the dissipation superoperator due to coupling to contact $K$. 
$\hat{\mathcal{D}}{}^{K\,(1)}_{e_1,e_2;e_3,e_4}(R,R',t;p,p',E,E')$ in Eq.(\ref{DK})
is the higher order correction to the dissipation superoperator. Its action on the Green function
$G^{<}_{e_3,e_4}(R',t;p',E')$ is 
\begin{align}
\label{DK1}
&\hat{\mathcal{D}}{}^{K\,(1)}_{e_1,e_2;e_3,e_4}(R,R',t;p,p',E,E')\, \mathbf{G}^{<}_{e_3,e_4}(R',t;p',E')
= 
\delta(R-R')\,\delta(p-p')\,\delta(E-E')\sum_{e_s,e_s'}\int dR_s\int dp_s\int \frac{dE_s}{2\pi}
\\ &
\frac{1}{2}\bigg(
\delta_{e_1,e_3}\big(\left\{\mathbf{G}^{<};c^{K>}-c^{K<}\right\}G^a 
               +\frac{\partial \mathbf{G}^{<}}{\partial E}\big[c^{K>}-c^{K<}\big]\frac{\partial G^a}{\partial t}\big)
+\delta_{e_2,e_4}\big(G^r\left\{c^{K<}-c^{K>};\mathbf{G}^<\right\}
                -\frac{\partial G^r}{\partial t}\big[c^{K<}-c^{K>}\big]\frac{\partial \mathbf{G}^{<}}{\partial E}\big)
 \bigg)
\nonumber  \\ &
 +\frac{1}{2}\sum_e\bigg(
 \big(\mathbf{G}^{<}\left\{c^{K>}-c^{K<};G^a\right\}
                 -\frac{\partial \mathbf{G}^{<}}{\partial t}\big[c^{K>}-c^{K<}\big]\frac{\partial G^a}{\partial E}\big)
+\big(\left\{G^r;c^{K<}-c^{K>}\right\}\mathbf{G}^{<}
                 +\frac{\partial G^r}{\partial E}\big[c^{K<}-c^{K>}\big]\frac{\partial \mathbf{G}^{<}}{\partial t}\big)
 \bigg)
 \nonumber 
\end{align}
\end{widetext}
where 
\begin{equation}
\{f_1;f_2\}\equiv \frac{\partial f_1}{\partial E}\frac{\partial f_2}{\partial t}
- \frac{\partial f_1}{\partial\vec p}\frac{\partial f_2}{\partial \vec R}
- \frac{\partial f_1}{\partial t}\frac{\partial f_2}{\partial E}
+  \frac{\partial f_1}{\partial\vec R}\frac{\partial f_2}{\partial \vec p}
\end{equation}
is the Poisson bracket. To shorten the notation we dropped the arguments in (\ref{DK1})
keeping in mind that the structure of the expression follows that of Eq.(\ref{DK0}). 
This correction is responsible for renormalizations of $G^{<}_{e_1,e_2}(R,t;p,E)$ 
similar to those discussed, e.g., in Ref.~\onlinecite{vonOppenPRB12}. 
In what follows we disregard this correction.
By doing so we get in Eq.(\ref{Gltgrad1}) usual structure of (energy and momentum resolved 
flavor of) quantum master equation and avoid complications related to consistency of the
gradient expansion procedure.\cite{BotermansMalflietPhysRep90}

\begin{widetext}
Performing first order gradient expansion in EOM (\ref{GrEOM}) leads to
\begin{equation}
\label{Grgrad1}
\sum_{e_3,e_4}\bigg(\hat{\mathcal{M}}{}_{e_1,e_2;e_3,e_4}(R,p,E)
-\sum_K\hat{\mathcal{S}}{}^{K}_{e_1,e_2;e_3,e_4}(R,t;p,E)\bigg) G^r_{e_3,e_4}(R,t;p,E)=\delta_{e_1,e_2}
\end{equation}
where
\begin{align}
&\hat{\mathcal{M}}{}_{e_1,e_2;e_3,e_4} (R,p,E)=
\delta_{e_1,e_3}\,\delta_{e_2,e_4}\bigg(E-\frac{p^2}{2}-V_{e_1}(R)\bigg)
\nonumber \\ &
-\delta_{e_2,e_4}\bigg(
f_{e_1,e_3}(R)+\bigg[i\vec p-\frac{1}{2}\frac{\partial}{\partial\vec R}\bigg]\vec d_{e_1,e_3}(R)
+\frac{1}{2}\frac{\partial}{\partial\vec R}\bigg[if_{e_1,e_3}(R)-\vec d_{e_1,e_3}(R)\cdot\vec p\bigg]
\frac{\partial}{\partial\vec p} + \frac{1}{2}\vec d_{e_1,e_3}(R)\frac{\partial}{\partial\vec R}
\bigg)
\\ &
-\delta_{e_1,e_3}\bigg(
f_{e_4,e_2}(R)-\bigg[i\vec p+\frac{1}{2}\frac{\partial}{\partial\vec R}\bigg]\vec d_{e_4,e_2}(R)
-\frac{1}{2}\frac{\partial}{\partial\vec R}\bigg[if_{e_4,e_2}(R)+\vec d_{e_4,e_2}(R)\cdot\vec p\bigg]
\frac{\partial}{\partial\vec p} + \frac{1}{2}\vec d_{e_4,e_2}(R)\frac{\partial}{\partial\vec R}
\bigg)
\nonumber
\end{align}
is the free propagation superoperator,\footnote{Such terms have been referred  to as mass superoperators\cite{IvanovKnollVoskresenskyNucPhysA00}} 
and
\begin{align}
\label{SK}
&\hat{\mathcal{S}}{}^{K}_{e_1,e_2;e_3,e_4}(R,t;p,E)=
  \hat{\mathcal{S}}{}^{K\,(0)}_{e_1,e_2;e_3,e_4}(R,t;p,E)+\hat{\mathcal{S}}{}^{K\,(1)}_{e_1,e_2;e_3,e_4}(R,t;p,E)
\\
\label{SK0}
& \hat{\mathcal{S}}{}^{K\,(0)}_{e_1,e_2;e_3,e_4}(R,t;p,E) =
\frac{i}{2}\sum_{e_s,e_s'}\int dR_s\int dp_s\int \frac{dE_s}{2\pi}\,
\nonumber \\ &
\bigg(
\delta_{e_2,e_4}G^r_{e_s,e_s'}(R_s,t;p_s,E_s)
\big(c^{K\,>}_{e_1,e_s;e_3,e_s'}(R,R_s,t;p,-p_s,E-E_s)
      -c^{K\,<}_{e_s',e_3;e_s,e_1}(R,R_s,t;-p,p_s,E_s-E)\big)
 \\ &
 +\delta_{e_1,e_3}G^r_{e_s',e_s}(R_s,t;p_s,E_s) 
 \big(c^{K\,>}_{e_4,e_s';e_2,e_s}(R,R_s,t;p,-p_s,E-E_s)
      -c^{K\,<}_{e_s,e_2;e_s',e_4}(R,R_s,t;-p,p_s,E_s-E\big)    
\bigg)
\nonumber
\end{align}
is the dissipation superoperator for the retarded Green function due to coupling to contact $K$. 
$\hat{\mathcal{S}}{}^{K\,(1)}_{e_1,e_2;e_3,e_4}(R,t;p,E)$ in Eq.(\ref{SK})
is the higher order correction to the dissipation superoperator. Its action on the Green function
$G^r_{e_3,e_4}(R,t;p,E)$ is
\begin{align}
\label{SK1}
&\hat{\mathcal{S}}{}^{K\,(1)}_{e_1,e_2;e_3,e_4}(R,t;p,E)\, \mathbf{G}^r_{e_3,e_4}(R,t;p,E) =
\frac{1}{4}\sum_{e_s,e_s'}\int dR_s\int dp_s\int \frac{dE_s}{2\pi}\,
 \\ &
\bigg(
\delta_{e_2,e_4}\big(-G^r\left\{c^{K<}-c^{K<};\mathbf{G}^r\right\}
      +\frac{\partial G^r}{\partial t}\big[c^{K>}-c^{K<}\big]\frac{\partial\mathbf{G}^r}{\partial E}\big)
+\delta_{e_1,e_3}\big(G^r\left\{c^{K>}-c^{K<};\mathbf{G}^r\right\}
      -\frac{\partial G^r}{\partial t}\big[c^{K>}-c^{K<}\big]\frac{\partial\mathbf{G}^r}{\partial E}\big)
\bigg)
\nonumber
\end{align}
\end{widetext}
The structure of Eq.(\ref{SK1}) follows that of (\ref{SK0}), 
which allows to reproduce the omitted indices.
In what follows we disregard this correction to the dissipation matrix. 
Eqs.~(\ref{Gltgrad1}) and (\ref{Grgrad1}) are the general final results of this paper. 
Next we turn to a specific simple example.
\section{Shifted harmonic oscillator}\label{osc}
In order to demonstrate relation to previous work we now consider a simple model of 
molecule represented by single level linearly coupled to a single harmonic oscillator.
There are only two electronic states in this problem, $\lvert 0\rangle$ and $\vert 1\rangle$,
corresponding to empty and occupied level, respectively. 
The matrix representing the molecular Hamiltonian, Eq.(\ref{HMe1e2}), becomes in this case
\begin{equation}
H^{M}_{e_1,e_2}(R)=\delta_{e_1,e_2}\bigg(-\frac{1}{2}\frac{\partial^2}{\partial R^2}+U_{e_1}(R)\bigg)
\end{equation}
where 
\begin{equation}
 U_e(R)=\frac{R^2}{2}+\delta_{e,1}(\varepsilon+\lambda R)
\end{equation}
Here $\varepsilon$ is position of the electronic level and $\lambda$ characterizes
the strength of coupling between the electron and molecular vibration (harmonic oscillator).

We are interested in energy resolved joint probabilities to observe the oscillator
at point $R$ with momentum $p$ while the electron level is empty, $P_0(R,t;p,E)$, 
or occupied,  $P_1(R,t;p,E)$. These probabilities are defined as
\begin{align}
P_0(R,t;p,E)\equiv& +i G^{<}_{00}(R,t;p,E)
\\
P_1(R,t;p,E)\equiv& -i G^{<}_{11}(R,t;p,E)
\end{align}
Then 
we get from Eq.(\ref{Gltgrad1})
\begin{widetext}
\begin{align}
\label{P0EOM}
&\bigg(\frac{\partial}{\partial t}+p\frac{\partial}{\partial R}-R\frac{\partial}{\partial p}\bigg)
P_0(R,t;p,E_0) =
\sum_K \int\frac{dE_1}{2\pi}\,\Gamma_K
\\ &\qquad\times
\bigg([1-f_K(E_{10})] A_{0}(R;p,E_0) P_1(R,t;p,E_1)
-f_K(E_{10}) A_{1}(R;p,E_1) P_0(R,t;p,E_0)\bigg)
\nonumber \\
\label{P1EOM}
&\bigg(\frac{\partial}{\partial t}+p\frac{\partial}{\partial R}-(R+\lambda)\frac{\partial}{\partial p}\bigg)
P_1(R,t;p,E_1) =
\sum_K\int\frac{dE_0}{2\pi}\,\Gamma_K
\\ &\qquad\times
\bigg(f_K(E_{10}) A_{1}(R;p,E_1) P_0(R,t;p,E_0)
-[1-f_K(E_{10})] A_{0}(R;p,E_0) P_1(R,t;p,E_1)\bigg)
\nonumber 
\end{align}
\end{widetext}
Here $E_{10}\equiv E_1-E_0$, $f_K(E)=[e^{\beta_K(E-\mu_K)}+1]^{-1}$ is 
the Fermi-Dirac thermal distribution in contact $K$,
$\Gamma_K\equiv2\pi\sum_{k\in K}\lvert V_k\rvert^2\delta(E-\varepsilon_k)$
is the electron escape rate to contact $K$ (wide band approximation is assumed),
and $A_e(R;p,E)=-2\,\mbox{Im}\, G_{ee}^r(R;p,E)$ is the many-body spectral function of the system,
Eq.(\ref{defA}).

Similarly, from Eq.(\ref{Grgrad1}) we get (e=0,1)
\begin{equation}
\label{Gree}
 G^r_{ee}(R;p,E)=\bigg[E-\frac{p^2}{2}-U_e(R)-\Sigma_{ee}^r(R;p,E)\bigg]^{-1}
\end{equation}
with
\begin{align}
&\Sigma^r_{00}(R;p,E_0)=
\\&\qquad -\frac{i}{2}\sum_K\Gamma_K\int\frac{dE_1}{2\pi}
f_K(E_{10}) G^r_{11}(R;p,E_1)
\nonumber
\end{align}
\begin{align}
&\Sigma^r_{11}(R;p,E_1)=
\\ &\qquad -\frac{i}{2}\sum_K\Gamma_K\int\frac{dE_0}{2\pi}
[1-f_K(E_{10})] G^r_{00}(R;p,E_0)
\nonumber
\end{align}

Eqs.~(\ref{P0EOM})-(\ref{Gree}) are the final results of this section. 
In spite of the simplification imparted by the gradient expansion their numerical 
solution presents a difficult task. Further simplifications are achieved in two limits:\\
(a) In the quasi particle approximation
($\Gamma\to 0$), when $\Sigma^r_{ee}\to -i0^{+}$,
we have from (\ref{defA}) and (\ref{Gree})
\begin{align}
 A_e(R;p,E)\equiv& -2\,\mbox{Im}\, G^r_{ee}(R;p,E)
 \\ &
 \overset{\Gamma\to 0}{\longrightarrow} 2\pi\delta\big(E-p^2/2-U_e(R)\big)
 \nonumber 
\end{align}
In this limit  expressions (\ref{P0EOM}) and (\ref{P1EOM})
reduce to those discussed in Ref.~\onlinecite{NitzanSubotnikJCP15}.\\
(b) When the molecule metal coupling is strong the electron exchange is fast relative 
to the characteristic nuclear  dynamics. In this case individual molecular electronic states cannot be probed and only the some of their probabilities is meaningful. 
In this case information on the different charging states becomes redundant, and
summing Eqs.~(\ref{P0EOM}) and (\ref{P1EOM}) we recover the Ehrenfest dynamics.\footnote{Note that in the simple model of Section~\ref{osc} the Ehrenfest dynamics is recovered for any value of $\Gamma$ summing (\ref{P0EOM}) and (\ref{P1EOM}). 
More general situation, Eq.~(\ref{Gltgrad1}), explicitly requires strong molecule-contacts 
coupling to recover the Ehrenfest dynamics.} 
\section{Conclusions}\label{conclude}
We have presented derivation of expressions for non-adiabatic molecular dynamics in junctions
starting from the full quantum-mechanical problem. The derivation starts from the exact EOMs
for the pseudo particle Green functions, describing junction's response in the language of 
many-body (vibronic) states of isolated molecule. Gradient expansion effectively separates
classical nuclear from quantum electron dynamics, yielding a Fokker-Planck equation which 
incorporates both optical (intra-molecular) and charge-transfer electron transitions 
as sources of non-adiabatic dynamics in junctions. The resulting equation can be 
viewed as the precursor of the surface-hopping algorithm. Indeed the surface hopping procedure described in Refs.~\cite{SubotnikJCP15,NitzanSubotnikJCP15_2,NitzanSubotnikJCP15_3,NitzanSubotnikJCP15} is obtained in the limit where level broadening 
is disregarded.
We also show that tracing out information on adiabatic surfaces leads to Ehrenfest
dynamics (motion on the potential of mean force).
Our study extends previous consideration by accounting for molecular hybridization with
contacts, and by introducing formulation capable of implementing
surface-hopping algorithm in a current carrying (bias induced) molecular junction. 
At equilibrium and in the limit of weak molecule-contacts coupling our results reduce
to those of Ref.~\onlinecite{NitzanSubotnikJCP15}.
Development of numerical codes capable of implementing the scheme is a complicated 
technical problem that is left for future effort.

\begin{acknowledgments}
We thank Michael Thoss and Philipp Werner for helpful discussions.
MG gratefully acknowledges support by the Department of Energy 
(Early Career Award, DE-SC0006422). 
The Research of AN is supported by the Israel Science Foundation and 
by the US-Israel Binational Science Foundation.
\end{acknowledgments}


\end{document}